\documentclass[aps,twocolumn]{revtex4}
\usepackage{graphicx}

\begin{document}

\title{Ultracold fermions and the SU($N$) Hubbard model}
\author{Carsten Honerkamp$^{1}$ and Walter Hofstetter$^{1,2}$ }
\affiliation{$^{1}$Department of Physics, Massachusetts Institute of 
Technology, Cambridge MA 02139, USA \\
$^2$Lyman Laboratory, Harvard University, Cambridge, MA 02138, USA}  
\date{September 16, 2003}
\begin{abstract} 
We investigate the fermionic SU($N$) Hubbard model 
on the two-dimensional square lattice 
for weak to moderate interaction strengths using one-loop
renormalization group and mean-field methods.  
For the repulsive case $U>0$ at half filling and small $N$ the 
dominant tendency is towards breaking of the SU($N$) symmetry.  
For $N>6$ staggered flux order takes over as the dominant 
instability, 
in agreement with the large-$N$ limit.  
Away from half filling for $N=3$ 
the system rearranges the particle densities such that two flavors
remain half filled by cannibalizing the third flavor. In the 
attractive case and odd $N$ a full Fermi surface
coexists with a superconductor in the ground state.  
These results may be relevant to
future experiments with cold fermionic atoms in optical lattices. 
\end{abstract}

\pacs{}
\maketitle
\vskip1pc

\paragraph*{Introduction:}
After the celebrated observation of Bose-Einstein
condensation \cite{bec} ultracold atom systems receive growing
attention in the field of condensed matter physics. 
Recently, also quantum degenerate Fermi gases have been realized 
\cite{deMarco,truscott,fer3,fer4}, opening up the possibility to 
study phenomena such as BCS superfluidity in a new context. 
As a further important advance, optical lattices have been used 
to realize the transition between a bosonic superfluid and a Mott 
insulator \cite{greiner}. It has thus been demonstrated that cold atoms
systems can become a very flexible and clean laboratory for many
exciting phenomena from the purview of condensed matter or interacting
many particle systems. In particular, it has been suggested 
\cite{hofstetter} 
that cold fermions in optical lattices may help to
understand the notorious complexities of strongly correlated solid 
state
systems such as the cuprate high--temperature superconductors.

Besides the realization of phenomena that are known to exist in some
form in condensed matter systems, it is also interesting to ask 
whether
the degrees of freedom offered by cold atoms could give rise to 
states 
of matter that do not have obvious counterparts in the physics of
interacting electrons.  Typical electron systems, at least in the
first approximation, possess SU(2) spin rotational symmetry which can
be broken spontaneously, leading to magnetic phenomena such as ferro--
and anti--ferromagnetism.  For alkali atoms, the nuclear spin $I$ and 
electron spin $S$ are combined in a hyperfine state.  
Its total angular momentum $F$ can be different from $1/2$, 
and for each $F$ there are $2F+1$ hyperfine states differing by
their azimuthal quantum number $m_F$. E.g. for the fermionic 
$^{40}$K, 
the nuclear 
spin is $I=4$ and the lowest hyperfine multiplet (at weak fields) 
has $F=9/2$. In magnetic traps only a subset of these $2F+1$ states 
(the \emph{low--field--seekers}) can be trapped \cite{deMarco}, 
but this constraint can be avoided by using all--optical traps 
\cite{granade}. 

In fact, coexistence of the three hyperfine states 
$|F=9/2, m_{F}=-5/2,-7/2,-9/2\rangle$ of $^{40}K$ in an optical trap 
has already been realized, 
with tunable interactions due to Feshbach resonances between 
$m_{F}=-5/2/-9/2$ and $m_{F}=-7/2/-9/2$, respectively \cite{regal}.
A situation with strong attractive interaction between all three 
components can be realized e.g. for the spin polarized states with 
$m_{s}=1/2$ 
in $^{6}$Li where the triplet scattering length $a=-2160a_{0}$ 
is anomalously large \cite{abraham}. 

Optical lattices are created by a standing light wave leading
to a periodic potential for the atomic motion of the form
$V(x)=V_{0} \sum_{i} \cos^{2}(k x_{i})$
where $k$ is the wavevector of the laser, $i$ labels the spatial
coordinates and the lattice depth $V_{0}$ is usually measured in units of the 
atomic recoil energy $E_{R}=\hbar^{2} k^{2}/2m$. 
In the following we will consider the 2D case where
$i=1,2$. It has been shown \cite{boseth} that
the \emph{Hubbard model} with a local density--density interaction
provides an excellent description of the low--energy physics.
Here we are interested in a situation where fermionic atoms with
$N$ different spin states (``flavors'') $m$ are loaded into the
optical lattice.
We thus consider a Hubbard Hamiltonian
\begin{equation} 
H =  -t \sum_{m,\langle ij \rangle} 
    \left[c^\dagger_{i,m} c_{j,m} + c^\dagger_{j,m} c_{i,m} \right] + \frac{U}{2} \sum_i n_i^2 \, .  \label{hubb}
\end{equation} 
Here $n_i=\sum_{m} n_{i,m}$ is the total number
density of atoms on site $i$ which can be written in terms of 
creation and annihilation operators according to 
$n_{i,m} = c^\dagger_{i,m} c_{i,m}$.  
The interaction (second term in Eq. \ref{hubb}) 
is invariant under local U($N$) rotations of the $N$
flavors with different $m$.  The hopping term of the atoms between 
nearest neighbors $\langle ij \rangle$
reduces the invariance of the complete
Hamiltonian to a global U($N$) symmetry.  Stripping off
the overall U(1) phase factor, we arrive at the SU($N$) Hubbard model. 
Note that in the optical lattice the effective Hubbard parameters are given by 
$t=E_{R} \left(2/\sqrt{\pi}\right) \xi^{3} \exp\left(-2 \xi^{2}\right)$ 
and $U=E_{R} a_{s} k \sqrt{8/\pi} \xi^{3}$ where 
$\xi=\left(V_{0}/E_{R}\right)^{1/4}$ and $a_{s}$ is the $s$--wave 
atomic scattering length. 

The fermionic SU($N$) Hubbard model for 
$U>0$ on the two-dimensional (2D) square lattice was studied 
in the large-$N$ limit \cite{marston} in the early days of 
high--$T_c$ 
superconductivity, mainly as a controllable limit connected to the 
then 
physically relevant case $N=2$. A generalized SU($N$) model could 
describe 
orbitally degenerate electronic states in crystals, but it is likely 
that in these systems different overlaps between the orbitals 
pointing in distinct lattice directions break the SU($N$) 
invariance.   

\begin{figure}
\includegraphics[width=.44\textwidth]{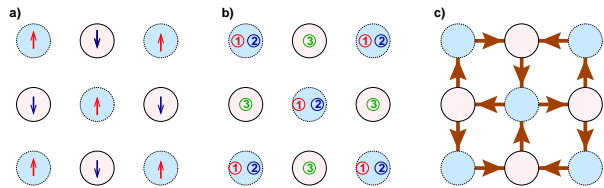}
\caption{a): AF spin-density wave state for $N=2$. Spin-up and 
spin-down particles occupy the two sublattices with different 
probabilities (here idealized to 0 and 1). b) SU(3)-breaking flavor-density wave state 
for $N=3$. Flavors 1 and 2 prefer one sublattice, flavor 3 the other. 
c) Staggered flux state for $N> 6$: the arrows indicate the particle 
currents.}
\label{orders}
\end{figure}

In the following we focus on the density region near half band 
filling with an average $N/2$ fermions per site. In the conventional 
$N=2$ Hubbard model at half filling the ground state exhibits 
spin-density wave (SDW) order, and when the filling is changed 
$d$-wave superconductivity is very likely \cite{zanchi}. The SDW
state breaks the translational invariance and spin-up and spin-down 
electrons (for staggered moment along the $z$-direction) 
occupy the two sublattices differently (see Fig. \ref{orders}). 
For large $N$ and small exchange interactions $J$, staggered flux order 
is expected to dominate over the SU($N$)-breaking 
states \cite{marston}.

\paragraph*{One-loop renormalization group for the half-filled band 
and general $N$ and $U>0$:} 
First let us analyze the one-loop renormalization group (RG) 
flow for the half filled band. 
We apply the perturbative temperature-flow RG 
method of Ref. \cite{tflow} that has proved to give good results for 
$N=2$. 
As initial condition one fixes the interaction at a high temperature 
of the order of the bandwidth. Then the RG flow describes the change 
of the interactions as the temperature is lowered and perturbative 
corrections due to one-loop particle-hole and particle-particle 
processes are taken into account. 
The interaction is described by a coupling function $V(\vec{k}_1,\vec{k}_2,\vec{k}_3)$ \cite{salmhofer}, where the flavor indices $m_1$ and $m_3$ belonging to the first incoming particle with wavevector $\vec{k}_1$ and the first outgoing particle with $\vec{k}_3$ are the same. 
Similarly $m_2=m_4$. The second outgoing wavevector $\vec{k}_4$ is fixed by momentum conservation on the lattice. 
As in the $N=2$ case the RG flow goes to strong coupling. This 
means, as we start the flow at high temperatures with a purely local 
interaction $V(\vec{k}_1,\vec{k}_2,\vec{k}_3)=U$, some coupling 
functions start to grow when the temperature is reduced and finally 
leave the perturbative range. 
At this temperature scale we stop the RG flow and analyze which class 
of coupling constants grows most strongly towards low $T$. In close 
analogy with the spin-1/2 case we consider couplings in the 
charge channel 
$ V_c(\vec{k},\vec{k}',\vec{q})= N 
V(\vec{k}+\vec{q},\vec{k}',\vec{k}) - V(\vec{k}', 
\vec{k}+\vec{q},\vec{k}) $
and in the SU($N$) symmetry breaking channel,
$ V_s(\vec{k},\vec{k}',\vec{q})= - V(\vec{k}', 
\vec{k}+\vec{q},\vec{k})$,
which, if divergent, signals a singular response for a small external 
field coupling to one of the $N^2-1$ generators of SU($N$).
We define averages over the Fermi surface,
$
\bar{V}^{\ell}_{c/s}(\vec{Q}) = \oint_{FS} d\phi_k 
\oint_{FS'} d\phi_{k'} \, g_\ell(k) g_\ell(k') V_{c/s} 
(\vec{k},\vec{k}',\vec{q}) \,
$
 with $g_1(\vec{k})=1$ in the $s$-wave channel or   $g_2(\vec{k})= 
(\cos k_x - \cos k_y) /\sqrt{2}$ in the $d$-wave channel.       
For $N=2$ the couplings $\bar{V}_s(\vec{Q})$ in the $s$-wave channel 
with 
$\vec{Q}=(\pi,\pi)$ diverge most strongly, 
indicating the formation of an antiferromagnetic (AF) SDW state. 
Note that for larger $N$ the charge couplings can become quite large 
as they scale with $N$. 

Analyzing the RG flows to strong coupling, the picture is as follows: 
for $N \le 6$ the $s$-wave coupling function $\bar{V}^s_s(\vec{Q})$ 
is the strongest divergent family of coupling constants, signalling a 
dominant tendency towards breaking of the SU($N$) symmetry with 
staggered two-sublattice real space dependence.  
For $N>2$ (especially for odd $N$) this leads to the interesting 
question how the $N/2$ particles per site will arrange themselves 
on the bipartite square lattice (see Fig. \ref{orders}).
Below we describe what happens in a mean-field analysis.

\begin{figure}
\includegraphics[width=.49\textwidth]{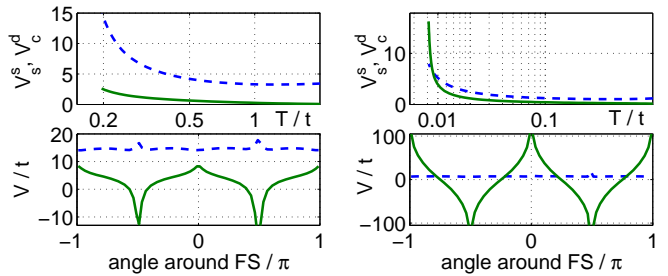}
  
\caption{
Left plots: RG results for SU(3) at half filling and $U=4t$. Upper left:
Flow of the coupling constants $\bar{V}_s^s$ in the SU(3)-breaking 
channel (dashed line)  and $\bar{V}_{c}^{d}$ in the staggered flux 
(SF) channel, averaged around the Fermi surface (FS). 
Lower left: 
$V_s(\vec{k},\vec{k}',\vec{Q})$ (dashed line) and 
$V_c(\vec{k},\vec{k}',\vec{Q})$ with $\vec{k}$ fixed on the FS at 
$(\pi,0)$ and $\vec{k}'$ moving around the FS after termination of the flow.
Right plots: The same for SU(8) at half filling. }
\label{s38plot}
\end{figure} 

For $N>6$ the flow to strong coupling changes qualitatively. Now the 
leading divergence is in the charge couplings $V_c 
(\vec{k},\vec{k}',\vec{Q})$ with a $d_{x^2-y^2}$-wave dependence 
on $\vec{k}$ and $\vec{k}'$. $\bar{V}_c^d$ diverges more strongly 
than $\bar{V}_s^s$ (see Fig. \ref{s38plot}), albeit at lower 
temperature $T \approx 0.014t$ for $U=4t$ and $N=7$. 
This signals a tendency towards
staggered flux (SF) order with long range ordering of the expectation 
value $\Phi_{SF} = \sum_{\vec{k},m} (\cos k_x -\cos k_y ) \langle 
c_{\vec{k},m}^\dagger c_{\vec{k}+\vec{Q},m} \rangle$.   
This result agrees with the large--$N$ limit
for small exchange interactions $J$ \cite{marston}.
 The SF state has surfaced several times for the SU(2) case in 
connection  with the high-$T_c$ cuprates and related 
models \cite{marston,sfrefs}, also as $d$-density wave state 
(although the particle density is {\em not} modulated). 
Its quasiparticles have a 
wavevector-dependent energy gap that vanishes at $\vec{k}=(\pm 
\pi/2,\pm \pi/2)$.
Nonzero $\Phi_{SF}$ breaks translational and time-reversal symmetry with 
alternating  particle currents around the plaquettes (see Fig. \ref{orders}). 
If the particles were charged, their motion would give rise to alternating 
magnetic moments pointing out of the plane, hence the name staggered 
flux state.  
Note that $\Phi_{SF}$ is SU($N$)-invariant and 
no continuous symmetries are broken.
Correspondingly the SF state can order at finite temperatures in 2D. 
For the same reason it may be possible that the staggered flux state 
sets in for somewhat lower $N$ than the critical $N=6$ in our 
one-loop RG study that neglects collective fluctuations.

Away from half filling  the flow of the dominant $(\pi,\pi)$- 
instability gets cut off at some low energy scale that increases 
with the distance to half filling.  
Below that scale there is a tendency towards  
$d_{x^2-y^2}$-wave Cooper pairing. However the energy scales for 
pairing instabilities become very small with increasing $N$. 
Below we discuss superfluid pairing for $N>2$ in the attractive 
case $U<0$.

\paragraph*{Ground state near half filling for $N=3$:} 
Having established that SU($N$) symmetry breaking at wave vector 
$(\pi,\pi)$ is the dominant instability of the repulsive model near half 
filling with $N<6$, we now turn to a mean-field description of the 
ground state for $N=3$. 
We decouple the interaction terms in the particle-hole channel with 
local mean-fields $\langle c_{\alpha,i}^\dagger c_{\beta,i}\rangle = 
M_{\alpha \beta, i}$.
The hermitian local mean-field matrix $M_{\alpha\beta}$
can be decomposed into a traceful part $M_0 $ proportional to the 
identity matrix $t_0$ and a traceless part $\sum_{a=1,\dots 8} M^a 
t_a$ with the 8 generators $t_a$ of the fundamental representation of 
SU(3). A finite value of one of the traceless components breaks the 
SU(3) invariance. We will now restrict the analysis to commensurate 
 order, where only uniform and staggered components of a 
commuting subset of the 9 $M^a$ acquire nonzero expectation values.  
SU(3) has rank 2 and the two diagonal generators commute mutually and 
with the identity matrix. We can choose these three degrees of 
freedom to be contained in the three flavor density  mean-fields 
$\langle n_{\alpha} \rangle$.

\begin{figure}
\includegraphics[width=.49\textwidth]{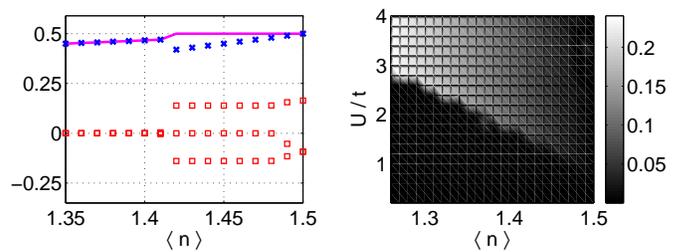}

\caption{Left: Uniform densities of the $N=3$ flavors (solid  
line: flavor 1 and 2, crosses: flavor 3) and staggered densities (squares)  vs. total 
density for $U=1.6t$. 
Right: Difference in uniform density between the two 
majority flavors and the minority flavor versus interaction $U$ and total density $\langle n \rangle$ per site. The scalebar 
indicates the density difference per site.}
\label{kplot}
\end{figure}

The results of $T=0$ mean-field solutions 
are shown in Fig. \ref{kplot}. At half filling, $n=1.5$/site, the 
SU(3)-breaking creates a flavor-density wave: 
two flavors  prefer one sublattice with equal density, 
while the third flavor goes  predominantly on the other sublattice with 
a somewhat larger density modulation. 
The staggered components 
do not add up to zero. Thus there is a charge density wave 
accompanying the SU(3) symmetry breaking. 
For $U=3t$ the mean-field $T_c$ for this state is $\sim 0.45t$, but in the one-loop RG is it is reduced down to $\sim 0.12t$.

Fig. \ref{kplot} also describes the results away from half filling.
For example at $U=1.6t$ and $1.42<n<1.48$/site, 
two flavors order with opposite staggered densities on the two 
sublattices, keeping their individual average density at half filling. 
Since the total density is less than half filling, 
the third flavor gets decimated with uniform density of $(n-1)$. 
As can be seen from the right plot in Fig. \ref{kplot}, this state 
only occurs above a critical 
interaction strength $U_c$ that increases from zero with increasing 
distance to $n=1.5$/site.
Note that the depletion of one flavor allows the system to preserve 
the commensurate order away from commensurate band filling.
A similar pinning of a part of the system 
to half filling is found in ladder systems \cite{rice}.
We add that for larger $U$ and close to half filling
($1.46< n <1.5$/site for $U=4t$) we find a another regime where the 
mean field equations converge slowly and microscopic phase separation 
might occur.

\paragraph*{Attractive SU($N$) Hubbard model:}
Next let us consider the attractive interactions $U<0$. 
In the SU(2) case in 2D, there is a power-law 
$s$-wave singlet superconductor/superfluid (SSC) below a 
Kosterlitz-Thouless 
transition  away from half filling \cite{scalettar}. At half filling 
the Kosterlitz-Thouless $T_c$ is zero, and SSC and 
charge-density wave (CDW) mean field states are degenerate and in the 
true ground state, both types of order coexist.
One-loop RG finds in this case that CDW and SSC susceptibilities are 
perfectly degenerate and diverge together at low $T$. This symmetry 
is destroyed for larger $N>2$ and the CDW susceptibility grows much 
faster than the one for SSC. Correspondingly we expect the ground 
state to have CDW long range order only. This is corroborated by a 
mean-field calculation for the SU(3) case that shows that the CDW 
order suppresses any kind of SSC admixture, and the CDW ground state 
energy is lower than that of the SSC state. 

We now consider the generic case sufficiently far away from half 
filling. Then the dominant instability is onsite
pairing. We decouple the interaction as
$ H_{U,\mathrm{mf}}= -\frac{1}{2} \sum_{\vec{k}, \alpha ,\beta}  
c^\dagger_{\vec{k}\alpha}   c^\dagger_{-\vec{k}\beta} \Delta_{\beta 
\alpha} + h.c. $
with the  local mean-fields $\Delta_{\alpha \beta} = -U \sum_{\vec{k}} 
\langle  
c_{\vec{k}\alpha}   c_{-\vec{k}\beta} \rangle = - \Delta_{\beta 
\alpha} $.
For $N>2$ these even parity gap functions 
$\Delta_{\alpha \beta}$ transform non-trivially under SU($N$). 
Depending on the global gauge, 
$\Delta_{\alpha \beta} $ takes different values. This is unlike 
the SU(2) case where even parity gap functions are singlets and 
invariant under spin rotations \cite{fn1}.
For SU(2) the ground state is degenerate with respect to the 
global phase of the gap function, and long-wavelength variations of 
the latter are gapless in absence of long-range forces. In the 
SU($N$) case we find a higher degeneracy and more gapless modes. It turns 
out that for SU(3) all gap functions with  
with  the same $\Delta_0^2= \sum_{\alpha \beta} |\Delta_{\alpha 
\beta}|^2$ are degenerate and have the same total density of states.
Apart from the global phase there are four additional gapless 
modes, 
two associated with the internal phases between $\Delta_{12}$, 
$\Delta_{13}$ and $\Delta_{23}$, and two modes modulating 
$|\Delta_{12}|$, 
$|\Delta_{13}|$ and $|\Delta_{23}|$ with fixed $\Delta_0$. 

A particularly simple choice in the degenerate manifold is 
$\Delta_{12}= \Delta_0$ and $\Delta_{13}=\Delta_{23}=0$. Then flavor 
3 remains completely unpaired and metallic. Since we can always 
rotate into this gauge, all SU(3) $s$-wave superconducting mean-field 
states are one-third (neutral) metals and two-thirds superfluids.
The gauge  with only $\Delta_{12}\not= 0$ makes the symmetry breaking 
pattern obvious. The original symmetry group of the problem SU(3) 
$\otimes$ U(1) with nine generators gets broken down to an SU(2) in 
flavor 1 and 2, leaving $\Delta_{12}$ invariant, and an additional 
U(1) that acts on the phase of the unpaired flavor 3. This leaves 5 
generators broken, yielding the collective modes described above.  
 For $3/8$ band filling and $U=4t$, the mean-field $T_c$ is $\sim 0.17t$.
The coexistence of a full Fermi surface with a superconductor should 
have interesting consequences. For example the collective modes may 
be subject to damping below twice the gap frequency and 
could hence render the ungapped fermionic spectrum observable.  
Experimentally, these Goldstone modes can be detected in the 
spectrum of elementary excitations measured via Bragg scattering 
off two non--collinear laser beams with frequency and momentum 
difference $\omega$,$q$ \cite{stenger}. 
By monitoring the number of scattered atoms as a function 
of $\omega$, this technique yields the dynamical 
structure factor $S(q,\omega)$. 
The collective modes will then lead to peaks in the scattering cross section. 

Theoretically, an additional weak $p$-wave attraction could trigger 
a superfluid transition of the unpaired flavor at much lower 
temperatures, leading to a coexistence of even- and odd-parity 
superfluidity.

The SU(4) case is more complicated. There the degeneracy 
of the ground state is subject to more constraints than just constant 
$\Delta_0$. The mean-field solutions have 
$|\Delta_{12}|=|\Delta_{34}|$, 
$|\Delta_{13}|=|\Delta_{24}|$ and  $|\Delta_{14}|=|\Delta_{23}|$. 
The single particle spectrum is fully gapped.

\paragraph*{Conclusions:} 
The fermionic SU($N$) Hubbard model on the 2D square lattice can 
possibly be realized with ultracold atoms in an optical lattice. Its 
ground states may exhibit phenomena that do not occur right away 
in traditional solid state systems. For $U>0$ we find a staggered 
flux state for large $N>6$ at half band filling where the particles 
run around the plaquettes of the lattice in an alternating 
way. This state has a partially gapped excitation spectrum with nodes 
along the Brillouin zone diagonals, which may be detectable via 
the momentum distribution function. 
Near half filling for 
$N=3$ we find a redistribution of the particle densities where two of 
the three flavors remain half filled and occupy different sublattices 
while the third flavor becomes depleted. Finally, in the attractive 
case $U<0$  we point out that the $s$-wave paired superfluid states 
may exhibit new collective modes. 
For $N=3$ a third of the particles remains 
ungapped, leading to a full Fermi surface coexisting with the 
superfluid. We expect this to be a general feature for odd $N$, also 
in three dimensions or in absence of a lattice potential.

Finally, a comment on the temperature scales 
which we have given in terms of the hopping parameter $t$. 
It has been shown \cite{hofstetter} that if the optical lattice 
is switched on slowly after termination of evaporative cooling, 
an additional \emph{adiabatic cooling} process takes place. 
The final temperature is given by the identity  
$T_{\rm initial}/T_{ F,\rm free} \approx T_{\rm final}/T_{ F,\rm lattice}$ 
where $T_{ F,\rm free (lattice)}$ denote the 
Fermi temperature of the free atomic cloud and in the presence 
of the lattice, respectively. In particular, in 2D at half filling 
one has $T_{ F,\rm lattice} = 4 t$. As a result, the critical atomic 
temperatures which have to be reached \emph{before} 
the lattice is switched on can be obtained from our results via the substitution 
$t \to T_{F, \rm free}/4$. For the $s$-wave superfluid phase ($U<0$) 
and the flavor-density wave states ($U>0$) 
we therefore find transition temperatures of order 
$0.05 T_{F}$ which are within experimental reach. 

We thank P.A. Lee, W.V. Liu, T.M. Rice and C.H. Schunck for useful 
discussions. E. Demler and M.D. Lukin are acknowledged for suggesting the topic 
to one of us. C.H. and W.H. were supported by the German Research Foundation (DFG) and
W.H. by a Pappalardo fellowship.


\begin{thebibliography}{99}
\bibitem{fn1} Note that for SU(2) the product of two fundamental 
representations $2 \otimes 2= 1 \oplus 3$ contains a singlet, while 
for SU(3) $3 \otimes 3 = \bar{3} \oplus 6$. Thus the even parity gaps span 
a three-dimensional subspace. SU(3)-singlets can only be formed in 
the particle-hole channel (cp. pions).
\end{thebibliography}
\end{document}